\documentclass[12pt,onecolumn]{revtex4}
\begin{document}

\title{Kerr-de Sitter Spacetimes in Various Dimensions\\
and dS/CFT Correspondence}
\author{M. H. Dehghani}
\address{Physics Department and Biruni  Observatory,
         Shiraz University, Shiraz 71454, Iran}
\begin{abstract}
We consider the Kerr-de Sitter (Kerr-dS) black hole in various
dimensions. Introducing a counterterm, we show that the total
action of these spacetimes are finite. We compute the masses and
the angular momenta of Kerr-dS spaces with one rotational
parameter in four, five and seven dimensions. These conserved
charges are also computed for the case of Kerr-dS space with two
rotational parameters in five dimensions. Although the angular
momentum density due to the counterterm is nonzero, it gives a
vanishing contribution to the total angular momentum. We also find
that the total angular momentum of the spacetime is independent of
the radius of the boundary for all cases, a fact that is not true
for the total mass of the system.
\end{abstract}

\maketitle

%\pacs{04.60.-m,04.62.+v,04.70.-s}
%\onecolumn

\section{Introduction\label{Intro}}

Spacetimes which are asymptotically de Sitter (dS) have attracted
a great deal of attention recently
\cite{Stro1,Bal1,Wit1,Kl1,Bal2,Stro2,Kl2}. First, this is due to
the recent astrophysical data indicating a positive cosmological
constant \cite{Perl}, and second because of the understanding of
the role of de Sitter spacetimes in string theory and
clarification of the microscopic origin of the entropy of dS
spacetimes. For AdS spacetimes an anti-de Sitter conformal field
theory (AdS/CFT) correspondence is known which posits a
relationship between supergravity or string theory in bulk AdS
spacetimes and conformal field theory on the boundary
\cite{Wit2,Haw1, Mal1,Deh2}. But, for de Sitter spaces, such a
relation between the bulk dS and conformal field theory on the
boundary is not yet known \cite{Kl2,Bal2}.

The concepts of action and energy momentum play central roles in
gravity, but, as is known, there is no good local notion of energy
for a gravitating system. A quasilocal definition of the energy
and conserved quantities can be found in \cite{BY}. The known
obstacle to the straightforward definition of the gravitational
action and therefore the conserved quantities of a gravitating
system is that the action and therefore all the conserved
quantities diverge. One approach toward evaluating them has been
to carry out all computations relative to some other spacetime
which is regarded as the ground state for the class of spacetimes
of interest. \ This is done by taking the original action
$I_G=I_v+I_b$\ for gravity fields and subtracting from it a
reference action $I_0$, which is a functional of the induced
metric $\gamma $ on the boundary $\partial \mathcal{M}$. Conserved
quantities are then computed relative to this boundary, which can
then be taken to (spatial) infinity if desired.

 This approach has been widely successful in providing the
conserved quantities of the system in regions of both finite and
infinite spatial extent \cite{BY,BCM}. Unfortunately, it suffers
from several drawbacks. The choice of reference spacetime is not
always unique \cite{CCM}, nor is it always possible to embed a
boundary with a given induced metric into the reference
background. Indeed, for Kerr spacetimes this latter problem forms
a serious obstruction towards calculating the subtraction energy,
and calculations have only been performed in the slow-rotating
regime \cite {Martinez}. An extension of this approach was
developed for asymptotically AdS spacetimes based on the
conjectured AdS /CFT correspondence
\cite{hensken,hyun,balakraus,Deh1}. Since quantum field theories
in general contain counterterms, it is natural from the AdS/CFT
viewpoint to append a boundary term $I_{ct}$ to the action which
depends on the intrinsic geometry of the (timelike) boundary at
large spatial distances. This requirement, along with general
covariance, implies that these terms should be functionals of
curvature invariants of the induced metric and have no dependence
on the extrinsic curvature of the boundary. An algorithmic
procedure \cite{kls} exists for constructing $I_{ct}$ for
asymptotically AdS spacetimes, and so its determination is unique.
Addition of $I_{ct}$\ will not affect the bulk equations of
motion, thereby eliminating the need to embed the given geometry
in a reference spacetime. Hence conserved quantities can now be
calculated intrinsically for any given spacetime. The efficacy of
this approach has been demonstrated in a broad range of examples,
all of them in the spatially infinite limit, where the AdS/CFT
correspondence applies \cite{nutent,EJM,nutkerr,dasman,Awad,vish}.

Recently, these ideas have been extended to the case of
asymptotically de Sitter spacetimes \cite{Kl1,Bal2}. The purpose
of this paper is to investigate the effects of including $I_{ct}$
for Kerr-dS spacetimes in four and higher dimensions. There are
several reasons for considering this. Although untill now there is
no well defined dS/CFT correspondence, from a gravitational
viewpoint the inclusion of additional boundary functionals is not
uniquely dependent upon this correspondence and can be carried out
even if the dS/CFT correspondence is found to be invalid. It is
therefore of interest to explore the implications of $I_{ct}$ for
more spacetimes. Furthermore, adding the counterterm would change
the value of conserved quantities of a finite gravitating system,
and thus for consideration of a finite gravitating system the
inclusion of $I_{ct}$ is also important. In this paper we extend
the idea of using the counterterm introduced by \cite {Kl1,Bal2}
for more than five dimensions. We specifically introduce a
counterterm for the class of Kerr-dS black holes in four to seven
dimensions. With their lower degree of symmetry relative to
spherically symmetric black holes, these spacetimes allow for a
more detailed study of the consequences of including $I_{ct}$.

The outline of our paper is as follows. In Sec. II we review the
basic formalism and introduce the counterterm. Section III will be
devoted to the consideration of Kerr-dS metric with one rotational
parameter in various dimensions, and we compute the conserved
quantities of these metrics. The conserved charges of the general
Kerr-dS spacetimes with two rotational parameters will be computed
in the next section. We finish our paper with some concluding
remarks.

\section{Stress Tensor and Conserved Quantities\label{StrTen}}

The gravitational action of $(n+1)$-dimensional de Sitter spacetimes is the
sum of two terms. The first term is the Einstein-Hilbert volume (or bulk)
term with positive cosmological constant $\Lambda =n(n+1)/(2l^2)$,
\begin{equation}
I_v=-\frac 1{16\pi }\int_{\mathcal{M}}d^{n+1}x\sqrt{-g}\left( \mathcal{R}-%
\frac{2n(n-1)}{l^2}\right) ,  \label{Actgv}
\end{equation}
\ and the second term is the Gibbons-Hawking boundary term:
\begin{equation}
I_b=\frac 1{8\pi }\int_{\partial \mathcal{M}}d^nx\sqrt{-\gamma }\Theta
(\gamma )  \label{Actgb}
\end{equation}
which is chosen such that the variational principle is well
defined. The Euclidean manifold $\mathcal{M}$ has metric $g_{\mu
\nu }$, covariant
derivative $\nabla _\mu $, and time coordinate $\tau $ which foliates $%
\mathcal{M}$ into nonsingular hypersurfaces $\Sigma _\tau $ with
unit normal $u_\mu $ over a real line interval $\Upsilon $.
$\Theta $ is the
trace of the extrinsic curvature $\Theta ^{\mu \nu }$ of any boundary(ies) $%
\partial \mathcal{M}$ of the manifold $\mathcal{M}$, with induced metric(s) $%
\gamma _{ij}$. In general $I_v$ and $I_b$ of Eqs. (\ref{Actgv})
and (\ref {Actgb})are both divergent when evaluated on solutions,
as is the Hamiltonian, and other associated conserved quantities
\cite{BY,BCM}. \ Rather than eliminating these divergences by
incorporating a reference term in the spacetime \cite{BCM,ivan}, a
new term $I_{ct}$ is added to the action which is a functional
only of boundary curvature invariants. Although there may exist a
very large number of possible invariants, one can add in a given
dimension only a finite number of them. For an asymptotically AdS
spacetime, these can be determined by an algorithmic procedure
\cite{kls}. \ These counterterms can be generalized to the case of
asymptotically dS spacetimes as
\begin{eqnarray}
I_{ct} &=&\frac 1{8\pi }\int_{\partial \mathcal{M}_\infty }d^nx\sqrt{-\gamma
}\{\frac{n-1}l-\frac{lR}{2(n-2)}-\frac{l^3}{2(n-4)(n-2)^2}[%
R_{ab}R^{ab}-\frac n{4(n-1)}R^2]  \nonumber \\
&& +\frac{l^5}{(n-6)(n-4)(n-2)^3}[\frac{(3n-2)R}{4(n-1)}%
R_{ab}R^{ab}-\frac{n(n+2)}{16(n-1)^2}R^3-2R^{ac}R_{abcd}R^{bd}+  \nonumber \\
&& \frac{n-2}{2(n-1)}R^{ab}\nabla _a\nabla _bR-R^{ab}\nabla
^2R+\frac 1{2(n-1)}R\nabla ^2R]+...\}, \label{Actct}
\end{eqnarray}
where $R$, $R_{abcd}$ and $R_{ab}$ are the Ricci scalar and
Riemann and Ricci tensor of the boundary metric $\gamma _{ab}$.

Now in the following sections we study the implications of including the
counterterm for the class of Kerr-dS metrics in various dimensions. As we
will show in these cases for a suitable choice of the parameter $m$ we have
the inner, outer, and cosmological horizons with radius $r_{-}$, $r_{+}$, and $%
r_c$. The trace of the extrinsic curvature, $\Theta $ and the square root of
the determinant of the metric, $\sqrt{-\gamma },$ are real for $r_{+}<$ $r<$
$r_c$ and imaginary for $r>r_c$. We will show that the total action
\begin{equation}
I=I_v+I_b+I_{ct}  \label{totact}
\end{equation}
will be finite for Kerr-dS spacetime in various dimensions for the
boundary at infinity. Of course one should note that, since for
$r>r_c$ the counterterm $I_{ct}$ is imaginary, it is necessary to
divide $I_{ct}$ by $i$ and then add it to the gravitational action
$I_v+I_b$ to get the total action.

Under the variation of the metric, one obtains
\begin{equation}
\delta I=[\hbox{terms that vanish when the equations of motion hold}]^{\mu
\nu }\delta g_{\mu \nu }+\int_{\partial \mathcal{M}}d^3x(P^{ab}+Q^{ab})%
\delta \gamma _{ab}  \label{varact}
\end{equation}
where $P^{ab}$ is related to the variation of the boundary term:
\begin{equation}
P^{ab}=\frac 1{8\pi }(\Theta ^{ab}-\Theta \gamma ^{ab})  \label{stres1}
\end{equation}
\ and $Q^{ab}$ is due to the variation of the counterterm
(\ref{Actct}) given as
\begin{eqnarray}
Q^{ab} &=&\frac 1{8\pi }\{-\frac{n-1}l\gamma ^{ab}+\frac l{n-2}(R^{ab}-\frac
12R\gamma ^{ab})+\frac{l^3}{(n-4)(n-2)^2}[-\frac 12\gamma
^{ab}(R^{cd}R_{cd}-\frac n{4(n-1)}R^2  \nonumber \\
&&\frac{nR}{2(n-2)}R^{ab}+2R_{cd}R^{acbd}-\frac{n-2}{2(n-1)}\nabla
^a\nabla ^bR+\nabla ^2R^{ab}-\frac 1{2(n-1)}\gamma ^{ab}\nabla
^2R]+...\}. \label{stres2}
\end{eqnarray}

One could decompose the metric $\gamma _{ab}$ in the form
\begin{equation}
\gamma _{ab}dx^adx^a=-N^2dt^2+\sigma _{ij}\left( d\phi ^i+V^i\right) \left(
d\phi ^j+V^j\right)   \label{gam}
\end{equation}
where the coordinates $\phi ^i$ are the angular variables
parametrizing the closed hypersurface of constant $r$ around the
origin. The conserved quantities associated with the stress
tensors of Eqs. (\ref{stres1}) and (\ref {stres2}) can be written
as
\begin{eqnarray}
\mathcal{Q}_1(\mathcal{\xi )} &=&\int_{\mathcal{B}}d^{n-1}\phi \sqrt{\sigma }%
P_{ab}n^a\mathcal{\xi }^b,  \nonumber \\
\mathcal{Q}_2(\mathcal{\xi )} &=&\int_{\mathcal{B}}d^{n-1}\phi \sqrt{\sigma }%
Q_{ab}n^a\mathcal{\xi }^b,  \label{cons}
\end{eqnarray}
where $\sigma $ is the determinant of the metric $\sigma =\gamma
_{ab}-n_an_b $, and $\mathcal{\xi}$ and $n^a$ is the Killing
vector field and the unit normal vector on the boundary
$\mathcal{T}$. As was mentioned both the conserved quantities of
Eqs. (\ref{cons}) are infinite, but a linear combination of them
is finite. Again, one should note that for the class of Kerr-dS
metrics in various dimensions the unit normal \ $n^a$ is timelike
for $r_{+}<$ $r<$ $r_c$ and spacelike for $r>r_c$.

For each Killing vector $\mathcal{\xi }$, there exists an
associated conserved charge. Thus for our case (Kerr-dS spacetimes
in various dimensions) with Killing vector ($\xi =\partial
/\partial t$) and rotational Killing vector ($\varsigma =\partial
/\partial \phi $)\ we obtain
\begin{eqnarray}
M &=&\int_{\mathcal{B}}d^{n-1}\phi \sqrt{\sigma
}(P_{ab}+Q_{ab})n^a\xi ^b,
\nonumber \\
J &=&\int_{\mathcal{B}}d^{n-1}\phi \sqrt{\sigma
}(P_{ab}+Q_{ab})n^a\varsigma ^b.  \label{charge}
\end{eqnarray}
These quantities are, respectively, the conserved mass and angular
momentum of the system enclosed by the boundary. Note that they
may both be dependent upon the location of the boundary
$\mathcal{B}$ in the spacetime, although each is independent of
the particular choice of foliation $\mathcal{B}$ within the
surface $\mathcal{T}$. We find that the different components of
the angular momentum are independent of the radius of the boundary
for the class of Kerr-dS spaces.

In the context of the dS/CFT correspondence the limit in which the boundary $%
\mathcal{B}$ becomes infinite is taken, and the counterterm
prescription ensures that the total action (\ref{totact}) and the
conserved charges (\ref {charge}) are finite. No embedding of the
surface $\mathcal{T}$ \ into a reference spacetime is required.
This is of particular advantage for the class of Kerr spacetimes,
in which it is not possible to embed an arbitrary boundary surface
into a flat (or constant-curvature) spacetime \cite
{Martinez,Deh1}.

\section{Kerr-dS Metric With One Rotational Parameter\label{Kerr}}

In this section we consider the class of Kerr-dS family of solutions in
various dimensions with one rotational parameter, which can be written as
\begin{eqnarray}
ds^2 &=&-\frac{\Delta _r}{\rho ^2}(dt-\frac a\Xi \sin ^2\theta d\phi )^2+%
\frac{\rho ^2}{\Delta _r}dr^2+\frac{\rho ^2}{\Delta _\theta }d\theta ^2
\nonumber \\
&&\ +\frac{\Delta _\theta \sin ^2\theta }{\rho ^2}[adt-\frac{(r^2+a^2)}\Xi
d\phi ]^2+r^2\cos ^2\theta d\Omega _{n-3}  \label{met1a}
\end{eqnarray}
in $(n+1)$ dimensions, where
\begin{eqnarray}
\Delta _r &=&(r^2+a^2)(1-\frac{r^2}{l^2})-2mr^{4-n},  \nonumber \\
\Delta _\theta  &=&1+\frac{a^2}{l^2}\cos ^2\theta ,  \nonumber \\
\Xi  &=&1+\frac{a^2}{l^2},  \nonumber \\
\rho ^2 &=&r^2+a^2\cos ^2\theta .  \label{met1b}
\end{eqnarray}

\subsection{Kerr-dS$_4$ metric:}

\ The metric of Eqs. (\ref{met1a}) and (\ref{met1b}) for $n=3$ has
three horizons located at $r_{\pm }$ and $r_c$,\ provided the
parameter $m$ lies in the range $m_{1,crit}\leq m\leq m_{2,crit}$
where $m_{1,crit}$ and $m_{2,crit}$ are the two critical masses
given by
\begin{eqnarray}
m_{1,crit} &\equiv &\frac l{3\sqrt{6}}\sqrt{1+33\frac{a^2}{l^2}(1-\frac{a^2}{%
l^2})-\frac{a^6}{l^6}-(1-14\frac{a^2}{l^2}+\frac{a^4}{l^4})^{3/2},}
\nonumber \\
m_{2,crit} &\equiv &\frac l{3\sqrt{6}}\sqrt{1+33\frac{a^2}{l^2}(1-\frac{a^2}{%
l^2})-\frac{a^6}{l^6}+(1-14\frac{a^2}{l^2}+\frac{a^4}{l^4})^{3/2}.}
\label{mcrit}
\end{eqnarray}

It is worthwhile to mention that in the limit $l\rightarrow \infty $, $%
m_{1,crit}=a$, $m_{2,crit}\rightarrow \infty $, $r_{\pm }=m\pm
\sqrt{m^2-a^2} $, and $r_c\rightarrow \infty $ as one may expect.
For $m=0$ the metric (\ref {met1a}) and (\ref{met1b}) is that of
pure dS$_4$ spacetime (or flat spacetime if $l\rightarrow \infty
$), and for $a=0$ the metric is that of Schwarzschild-dS$_4$
spacetime which has zero angular momentum. Hence we expect the
parameters $m$ and $a$ to be associated with the mass and angular
momentum of the spacetime respectively. As Eqs. (\ref{mcrit}) show
the critical masses for a Schwarzschild black-hole are $0$ and
$l/3\sqrt{3}$ \cite{Kl1}.

Using Eqs. (\ref{totact}), one could show that the total action is finite.
The total mass $M$ and the total angular momentum $J_\phi $ calculated from
Eqs. (\ref{charge}) are given by

\begin{eqnarray*}
M &=&-\frac m\Xi , \\
J_\phi  &=&\frac{ma}{\Xi ^2}.
\end{eqnarray*}

\subsection{Kerr-dS$_5$ metric}

For the case of $n=4$ the metric given by Eqs. (\ref{met1a}) and
(\ref{met1b}) has an outer and a cosmological horizon provided the
parameter $m$ lies in the range $ m_{1,crit}\leq m\leq m_{2,crit}$
where $m_{1,crit}$ and $m_{2,crit}$ are the two critical masses
given by
\begin{eqnarray}
m_{1,crit} &\equiv &\frac 12a^2,  \nonumber \\
m_{2,crit} &\equiv &\frac{l^2}8\left(1+\frac{a^2}{l^2}\right).
\label{mcrit5}
\end{eqnarray}
For $m=0$ the metric (\ref{met1a}) and (\ref{met1b}) is that of
pure dS$_5$ spacetime (or flat spacetime if $l\rightarrow \infty
$), and for $a=0$ the metric is that of Schwarzschild-dS$_5$
spacetime which has zero angular momentum. \ In this case the
critical masses are $0$ and $l^2/8$ \cite{Kl1}. Again the
parameters $m$ and $a$ are associated with the mass and angular
momentum of the spacetime, respectively.

Using Eqs. (\ref{totact}), one can show that the total action is
finite. The total mass $M$ and the total angular momentum $J_\phi
$ calculated from Eqs. (\ref{charge}) are given by

\begin{eqnarray*}
M &=&-\frac{3\pi }4\frac m\Xi +\frac{\pi l^2}{96}\left[7+\frac
1\Xi (\Xi^2+1)\right],\\
J_\phi  &=&\frac \pi 2\frac{ma}{\Xi ^2}.
\end{eqnarray*}

\subsection{Kerr-dS$_7$ metric:}

The Kerr-dS$_7$ metric with one rotational parameter given by Eqs.
(\ref{met1a}) and (\ref{met1b}) has two horizons located at
$r_{+}$ and $r_c$,\ provided the parameter $m$ is sufficiently
large relative to the other parameters, specifically,
\begin{equation}
m\geq m_{crit}\equiv \frac{l^4}{27}\left\{1+\frac 32\frac{a^2}{l^2}(1-\frac{a^2}{%
l^2})-\frac{a^6}{l^6}+(1+\frac{a^2}{l^2}+\frac{a^4}{l^4})^{3/2}\right\}.
\label{mcrit5}
\end{equation}

For $m=0$ the metric (\ref{met1a} and \ref{met1b}) is that of pure dS$_7$
spacetime (or flat spacetime if $l\rightarrow \infty $), and for $a=0$ the
metric is that of Schwarzschild-dS$_7$ spacetime which has zero angular
momentum, and the critical mass is $2l^4/27$. \ Again parameters $m$ and $a$
should be associated with the mass and angular momentum of the spacetime
respectively.

Using Eqs. (\ref{totact}), one can show that the total action is finite.
Again the total mass $M$ and the total angular momentum $J_\phi $ can be
calculated from Eqs. (\ref{charge}) as:

\begin{eqnarray*}
M &=&-\frac{5\pi ^2}8\frac m\Xi +\frac{\pi
^2l^4}{1280}\left[50-(1-\frac 6\Xi
)(\Xi -1)^2\right], \\
J_\phi  &=&\frac{\pi ^2}4\frac{ma}{\Xi ^2}.
\end{eqnarray*}

Remarkably, we find that the counterterm $Q_{ab}n^a\varsigma ^b$,
while non-zero, gives a vanishing contribution upon integration in
Eq. (\ref{charge}) for various dimensions of Kerr-dS metrics
considered in this section. Also one should note that the total
angular momentum $J_\phi $ for all the above cases is independent
of the radius of the boundary. This was also true for the case of
Kerr-AdS spacetime discussed in \cite{Deh1}.

\section{The General Kerr-dS Metric in Five Dimension}

It is known that the number of independent rotational parameters
for a metric with rotation group $SO(n)$ is the integer part of
$n/2$. Thus, the metric of the Kerr-dS$_5$ spacetime can have at
most two rotational parameters. The metric can be written as
\begin{eqnarray}
ds^2 &=&-\frac{\Delta _r}{\rho ^2}\left(dt-\frac{a\sin ^2\theta }{\Xi _a}d\phi -%
\frac{a\cos ^2\theta }{\Xi _a}d\psi \right)^2+\frac{\rho ^2}{\Delta _\theta }%
d\theta ^2+\frac{\rho ^2}{\Delta _r}dr^2  \nonumber \\
&&+\frac{1-r^2/l^2}{r^2\rho ^2}\left[abdt-\frac{b(r^2+a^2)\sin ^2\theta }{\Xi _a}%
d\phi -\frac{a(r^2+b^2)\cos ^2\theta }{\Xi _b}d\psi \right]^2  \nonumber \\
&&+\frac{\Delta _\theta \sin ^2\theta }{\rho ^2}\left(adt-\frac{r^2+a^2}{\Xi _a}%
d\phi \right)+\frac{\Delta _\theta \cos ^2\theta }{\rho
^2}\left(bdt-\frac{r^2+b^2}{\Xi _b}d\phi \right),  \label{metr2a}
\end{eqnarray}
where
\begin{eqnarray}
\Delta _r &=&\frac
1{r^2}(r^2+a^2)(r^2+b^2)\left(1-\frac{r^2}{l^2}\right)-2m,
\nonumber
\\
\Delta _\theta  &=&1+\frac{a^2}{l^2}\cos ^2\theta +\frac{b^2}{l^2}\sin
^2\theta ,  \nonumber \\
\Xi _a &=&1+\frac{a^2}{l^2},\hspace{.5cm}\Xi _b=1+\frac{b^2}{l^2},  \nonumber
\\
\rho ^2 &=&r^2+a^2\cos ^2\theta +b^2\sin ^2\theta .  \label{metr2b}
\end{eqnarray}
Again this metric has two event and a cosmological horizon
provided the parameter $m$ lies between the two critical masses
given by Eqs. (\ref {mcrit5}). Using Eqs. (\ref{totact}) and Eqs.
(\ref{charge}) one can show that the total action is finite and
the total mass $M$ and the total angular momenta $J_\phi $ and
$J_\psi $ are
\begin{eqnarray*}
M &=&-\frac{3\pi m}{4\Xi _a\Xi _b}+\frac{\pi
l^2}{96}\left[7+\frac{(\Xi^2 _a+\Xi^2
_b)}{\Xi _a\Xi _b}\right], \\
J_\phi  &=&\frac \pi {2\Xi _a^2\Xi _b}ma,\hspace{.5cm }J_\psi =\frac \pi
{2\Xi _a\Xi _b^2}ma.
\end{eqnarray*}

Again it is worthwhile to mention that the $\phi $ and $\psi $
components of the angular momentum due to the counterterm is zero,
and the total angular momenta $J_\phi $ and $J_\psi $ are
independent of the radius of the boundary.

\section{Closing Remarks}

In this paper we have computed the conserved charges of Kerr-dS spacetimes
in various dimensions through the use of the Brown-York boundary stress
tensor. Of course this is the first step in studying the prospects for a
duality between quantum gravity on de Sitter space and a Euclidean field
theory defined on the boundary.

By introducing a counterterm, $I_{ct}$ we showed that the total
action will be finite as the boundary goes to infinity. Although,
the angular momentum densities due to the counterterm for various
dimensions are not zero, they give vanishing contributions upon
integration. Thus, the total angular momenta of the spacetimes for
various dimensions are due only to the boundary terms which are
independent of the radius of the boundary, a fact that is not true
for the total mass of the spacetime. As in the case of anti-de
Sitter spaces the total mass in odd dimensions has a term which is
proportional to the parameter $m$, and a second term which is due
to the de Sitter spaces.

\end{document}